\documentclass[runningheads]{llncs}
\usepackage{graphicx}
\usepackage{amssymb}
\usepackage{pifont}
\newcommand{\cmark}{\ding{51}}
\newcommand{\xmark}{\ding{55}}

\usepackage{colortbl}
\usepackage{makecell}

\begin{document}

\title{DeepMCAT: Large-Scale Deep Clustering for Medical Image Categorization\thanks{Final authenticated publication is available online at https://doi.org/10.1007/978-3-030-88210-5\_26}}
\titlerunning{DeepMCAT: Large-Scale Deep Clustering for Medical Image Categorization}
%
\author{Turkay Kart\inst{1} 
\and Wenjia Bai\inst{1,2}
\and Ben Glocker\inst{1}
\and Daniel Rueckert\inst{1,3}
}
%
\authorrunning{T. Kart et al.}
%
\institute{BioMedIA Group, Department of Computing, Imperial College London, UK  \and
Department of Brain Sciences, Imperial College London, UK \and
Institute for AI and Informatics in Medicine, Klinikum rechts der Isar, Technical University of Munich, Germany \\
\email{t.kart@imperial.ac.uk}}
\maketitle              
\begin{abstract}

In recent years, the research landscape of machine learning in medical imaging has changed drastically from supervised to semi-, weakly- or unsupervised methods. This is mainly due to the fact that ground-truth labels are time-consuming and expensive to obtain manually. Generating labels from patient metadata might be feasible but it suffers from user-originated errors which introduce biases. In this work, we propose an unsupervised approach for automatically clustering and categorizing large-scale medical image datasets, with a focus on cardiac MR images, and without using any labels. We investigated the end-to-end training using both class-balanced and imbalanced large-scale datasets. Our method was able to create clusters with high purity and achieved over 0.99 cluster purity on these datasets. The results demonstrate the potential of the proposed method for categorizing unstructured large medical databases, such as organizing clinical PACS systems in hospitals. 

\end{abstract}
\section{Introduction}

Highly curated labelled datasets have recently been emerging to train deep learning models for specific tasks in medical imaging. Thanks to these fully-annotated images, supervised training of convolutional neural networks (CNNs), either from scratch or by fine-tuning, has become a \textit{dominant} approach for automated biomedical image analysis. However, the data curation process is often manual and labor-intensive as well as requiring expert domain knowledge. This time-consuming procedure is simply not practical for each single task in medical imaging, and therefore, automation is a necessity.  

The first step of data curation in medical imaging typically starts from data cleaning where desired images are extracted from a hospital image database such as a PACS system. Due to the nature of such image databases in hospitals, these systems often record important attributes such as image sequences in an unstructured fashion as meta-data in the DICOM header of the images. Meta-data in the DICOM standard, the most widely adapted format for data storage in medical imaging, may seem as a reliable option for automated annotation but it is often incorrect, incomplete and inconsistent. This represents a major challenge for data curation. Gueld et. al. \cite{gueld2002dicomquality} analyzed the quality of the DICOM tag \textit{Body Part Examined} in 4 imaging modalities at Aachen University Hospital and found that, in 15\% of the cases, the wrong information had been entered for the tag because of the user-originated errors. Misra et. al. \cite{misra2016humanbias} reported that labelling with the user-defined meta-data containing inconsistent vocabulary may introduce human-reporting bias in datasets, which degrades the performance of deep learning models. Categorization can be even more difficult for images stored in other formats, e.g. NIfTI in neuroimaging,  where meta-data is limited and/or simply not available for image categorization.

To categorize medical images in a realistic scenario, designing fully supervised methods would require a prior knowledge about the data distribution of the entire database, accounting for long-tailed rare classes and finally devoting significant effort to accurately and consistently obtaining manual ground-truth. In this work, we propose a different paradigm by efficiently using abundant unlabelled data and perform unsupervised learning. Specifically, we demonstrate that large-scale datasets of cardiac magnetic resonance (CMR) images can be categorized with a generalizable clustering approach that uses basic deep neural network architectures. Our intuition is that categorization of unknown medical images can be achieved if clusters with high purity are generated from learned image features without any supervision. Our approach builds on a recent state-of-the-art method, DeepCluster \cite{caron2018deepcluster}. 

Our main contributions are the following: (i) we show that pure clusters for CMR images can be obtained with a deep clustering approach; (ii) we investigate end-to-end training of the approach for both class-balanced dataset and highly imbalanced data distributions, the latter being particularly relevant for medical imaging applications where diseases and abnormal cases can be rare; (iii) we discuss the design considerations and evaluation procedures to adapt deep clustering for medical image categorization. To the best of our knowledge, this is the first study to perform simultaneous representation learning and clustering for cardiac MR sequence/view categorization and evaluating its performance on a large-scale imbalanced dataset (n = 192,272 images).

\section{Related Work}

A number of self-supervised and unsupervised methodologies have been explored to train machine learning models with abundant unlabelled data. In self-supervised learning (SSL), a pretext task is defined to train a model without ground-truth. While several studies have been explored in the context of self-supervision \cite{gidaris2018unsupervised,bai2019self}, domain expertise is typically needed to formulate a pretext task unlike our work. Similar to self-supervised learning, different strategies of unsupervised learning have been implemented with generative networks \cite{donahue2016adversarial} and deep clustering \cite{yang2016jule} to learn visual features. In this study, we focus on unsupervised deep clustering approaches at large scale. Although this has been investigated in a number of studies for natural images \cite{caron2019unsupervised,caron2018deepcluster}, various attempts in medical imaging have explored them with only limited amount of curated data in contrast to our methodology. 

Moriya et. al. \cite{moriya2018unsupervised} extended the JULE framework \cite{yang2016jule} for simultaneously learning image features and cluster assignments on 3D patches of micro-computed tomography (micro-CT) images with a recurrent process. Perkonigg et. al. \cite{perkonigg2020unsupervised} utilized a deep convolutional autoencoder with clustering whose loss function is a sum of reconstruction loss and clustering loss to predict marker patterns of image patches. Ahn et. al. \cite{ahn2019unsupervised} implemented an ensemble method of deep clustering methods based on K-means clustering. Pathan et. al. \cite{pathan2020ynet} showed clustering can be improved iteratively with joint training for segmentation of dermoscopic images. Maicas et. al. \cite{maicas2020unsupervised} combined deep clustering with meta training for breast screening.

One related approach to our study is the "Looped Deep Pseudo-task Optimization" (LDPO) framework proposed by Wang et. al. \cite{wang2017unsupervised}. LDPO extracts image features with joint alternating optimization and refine clusters. It requires a pre-trained model (trained on medical or natural images) at the beginning to extract features from radiological images and then fine-tunes the model paramaters by joint learning. Therefore, the LDPO framework starts with a priori information and strong initial signal about input images. On the contrary, our model is completely unsupervised and trained from scratch with no additional processing. In addition, we do not utilize any stopping criteria, which is another difference from LPDO \cite{wang2017unsupervised}. 

\section{Method}

Our method builds upon the framework of DeepCluster \cite{caron2018deepcluster}. The idea behind their approach is that a CNN with random parameters $\theta$ provides a weak signal about image features to train a fully-connected classifier reaching an accuracy (12\%) higher than the chance (0.1\%) \cite{noroozi2016jigsaw}. DeepCluster \cite{caron2018deepcluster} combines CNN architectures and clustering approaches, and it proposes a joint learning procedure. The joint training alternates between extracting image features by the CNN and generating pseudo-labels by clustering the learned features. It optimizes the following objective function for a training set $X = \{x_1, x_2, ..., x_N\}$:

\begin{equation} 
\min_{\theta, W} \frac{1}{N} \sum_{n=1}^{N} \ell(g_w(f_{\theta}(x_n)), y_n)
\end{equation} 

\noindent Here $g_w$ denotes a classifier parametrized by $w$, $f_{\theta}(x_n)$ denotes the features extracted from image $x_n$,  $y_n$ denotes the pseudo-label for this image and $l$ denotes the multinominal logistic loss \cite{caron2018deepcluster}. Pseudo-labels are updated with new cluster assignments at every epoch. To avoid trivial solutions where output of the CNN is always same, the images are uniformly sampled to balance the distribution of the pseudo-labels \cite{caron2018deepcluster}. 

\begin{figure}
\centering
\includegraphics[width=0.8\textwidth]{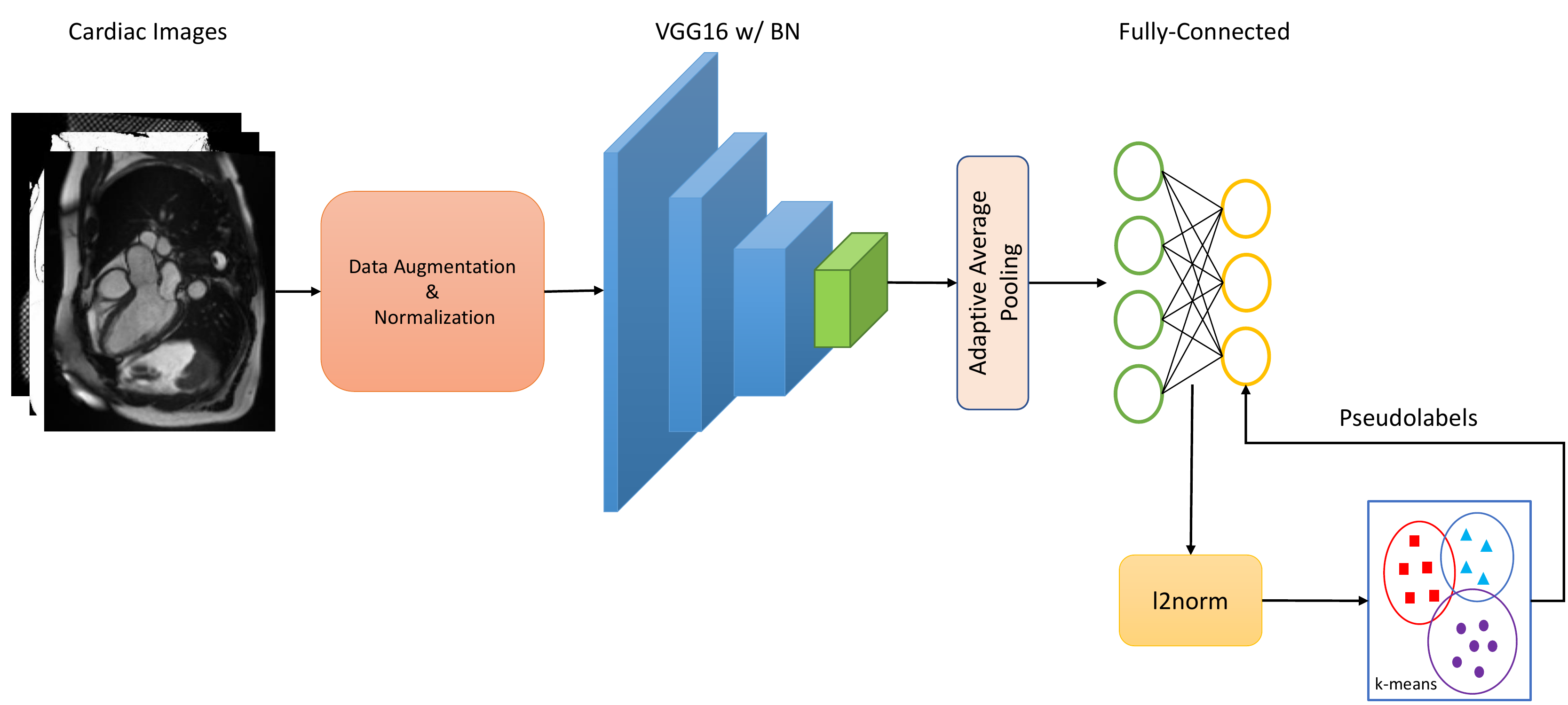}
\caption{Entire processing pipeline of our method based on DeepCluster \cite{caron2018deepcluster}} \label{figure}
\end{figure}

In this study, we keep parts of DeepCluster \cite{caron2018deepcluster} such as VGG-16 with batch normalization \cite{simonyan2014vgg} as the deep neural architecture and K-means \cite{johnson2019billion} as the clustering method, and then we adapt the rest for cardiac MR image categorization, illustrated in Fig.~\ref{figure}. To begin with, we add an adaptive average pooling layer between the VGG's last feature layer and the classifier. In DeepCluster \cite{caron2018deepcluster}, PCA is performed for dimensionality reduction which results in 256 dimensions whereas we preserve the original features. These features are $\ell_{2}$-normalized before clustering. DeepCluster \cite{caron2018deepcluster} feeds Sobel-filtered images to the CNN instead of raw images. In contrast, our method uses raw cardiac MR images in our experiments. We utilize heavy data augmentations including random rotation, resizing and cropping with random scale/aspect ratio for both training and clustering. Lastly, we normalize our images with z-scoring independently instead of using global mean and standard deviation.

We utilize the UK Biobank cardiac MR dataset which is open to researchers and contains tens of thousands of subjects. The whole dataset contains 13 image sequences/views, including short-axis (SA) cine, long-axis (LA) cine (2/3/4 chamber views), flow, SHMOLLI, etc \cite{petersen2015uk}. These images are in 2D, 2D + time or 3D + time. UK Biobank employs a consistent naming convention for different cardiac sequences and view-planes. We generated ground-truth labels using this naming convention and classified images into 13 categories \cite{bai2018cardiovascular}. To investigate the effect of class distribution on our methodology as well as the training stability, we designed three experiment settings using subsets of the entire dataset: (i) a subset of 3 well-balanced classes (LA 2 / 3 / 4 chamber views), and (ii) the large dataset of and (iii) the smaller dataset of high class imbalance of 13 classes. In these datasets, 2D images at t=0 were saved in PNG format for faster loading and training. If the images are in 3D + time, every single slice in z direction at t=0 were saved. Total numbers were 47,637 images in the dataset (i), 192,272 images in the dataset (ii), 23,943 images in dataset (iii). Example images are illustrated in Fig. 4, and the class distributions are reported at the Table 2 in the supplementary material.

\section{Results and Discussion}

In our experiments, we followed a systematic analysis of the proposed methodology. We want to answer these four questions below:
    \begin{enumerate}
        \item Is it feasible to categorize uncurated large-scale cardiac MR images based on their cluster assignments?
        \item How does the class balance affect deep clustering for medical images?
        \item How stable is training given there are no clear stopping criterion?
        \item How should we interpret the evaluation metrics?
    \end{enumerate}

\subsubsection{Experiment settings:} For training, we set the total number of epochs as 200. Our optimizer was stochastic gradient descent (SGD) with momentum 0.9 and weight decay of 1e-5. Our batch size was 256 and initial learning rate was 0.05. In the literature, there is a large body of empirical evidence which indicates that over-segmentation improves the performance of a deep clustering method \cite{caron2018deepcluster}. Based on this evidence, we set the number of clusters to be 8 times of number of classes in the datasets, which corresponded to 24 for the dataset of 3 well-balanced classes, and 104 for the datasets with 13 classes. 

\subsubsection{Evaluation metrics:} We used normalized mutual information (NMI) \cite{schutze2008introduction} and cluster purity (CP) \cite{schutze2008introduction} to evaluate the clustering quality of our models. 

\begin{equation}
    NMI(X, Y) = \frac{2I(X;Y)}{H(X) + H(Y)}
\end{equation}
Here $I$ is the mutual information between $X$ and $Y$ and $H$ is the entropy. For our experiments, we calculate two NMI values: NMI against the previous cluster assignments ($t-1$) and NMI against ground-truth labels.

\begin{equation}
    CP(X, L) = \frac{1}{N}\displaystyle\sum\limits_{k}\max_{j}|{x_k\cap l_j|}
\end{equation}
Here $N$ is the number of images, $X$ are the cluster assignments at epoch $t$ and $L$ is the ground-truth labels.

Accurate interpretation of our metrics, CP and NMI, is important. CP has a range from 0 to 1, which shows poor and perfect clusters, respectively. As the number of clusters increases, CP generally tends to increase until every image forms a single cluster, which achieves perfect clusters. In addition, we utilize NMI which signifies the mutual shared information between cluster assignments and labels. If clustering is irrespective of classes, i.e. random assignments, NMI has a value of 0. On the other hand, if we can form classes directly from cluster assignments, then NMI has a value of 1. The number of clusters also affects the NMI value but normalization enables the clustering comparison \cite{schutze2008introduction}. In our experiments, we did not employ any stopping criteria; thus, we always used the last model. In addition, during the training, we did not use NMI between cluster assignments and ground-truth, or cluster purity for validation.

\setlength{\tabcolsep}{0.305em} 
{\renewcommand{\arraystretch}{1.2}
\begin{table}[t]
\centering
\arrayrulecolor{black}
\caption{Performance of our method for different data configurations after 200 epochs}
\begin{tabular}{cccccccc}
\arrayrulecolor{black}\cline{1-8}
\thead{Dataset} & \thead{Balanced \\ Classes} & \thead{\# of \\ Images} & \thead{\# of \\ Classes} & \thead{\# of \\ Clusters} & \thead{NMI \\ t vs t-1} & \thead{NMI \\ t vs labels} & \thead{Cluster \\ Purity}  \\ 
\arrayrulecolor{black}\cline{1-8}
(i) & \cmark & 47,637 & 3  & 24 & 0.675 & 0.519 & 0.997  \\
(ii) & \xmark & 192,272 & 13 & 104 & 0.782 & 0.605 & 0.991  \\
(iii) & \xmark & 23,943 & 13 & 104 & 0.745 & 0.609 & 0.994 
\end{tabular}
\label{results}
\end{table}
}

\subsubsection{Discussion:} 
\begin{figure}
\centering
\includegraphics[width=\textwidth]{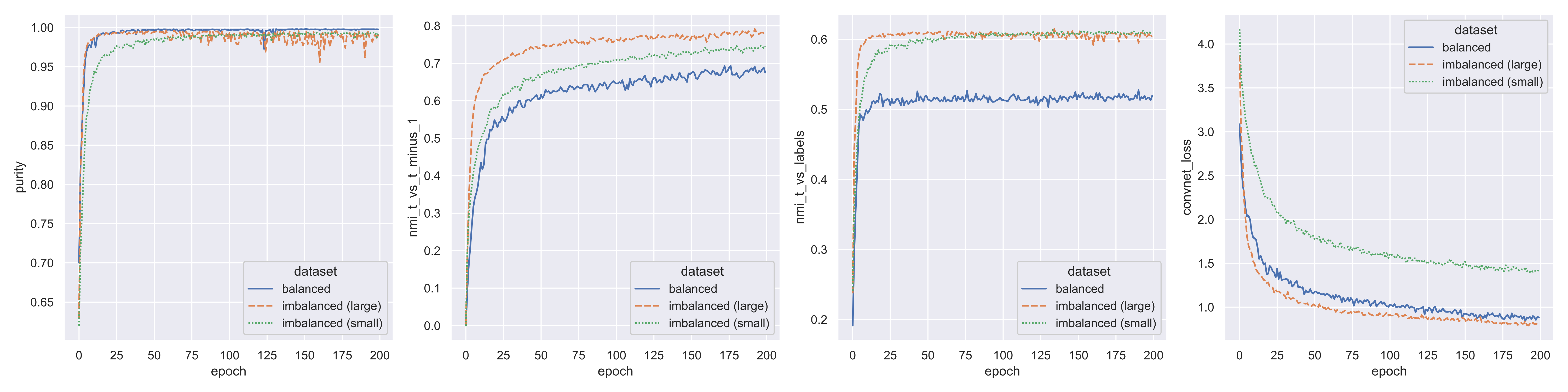}
\caption{Training metrics of our method for different data configurations} \label{metrics_2}
\end{figure}

Metrics and loss progression throughout the training are given at Fig.~\ref{metrics_2}. Results of our deep clustering method, which are calculated from features at the 200th epoch, are given at Table~\ref{results}. Our method is able to reach a clustering purity above 0.99 for both class balanced and imbalanced datasets, which shows the feasibility of the deep clustering pipeline to categorize large-scale medical images without any supervision or labels. The class imbalance does not affect overall performance but balanced classes provide a more stable purity throughout the training. We also show that a relatively smaller dataset can be enough for efficient clustering with high cluster purity.

\begin{figure}
\includegraphics[width=\textwidth]{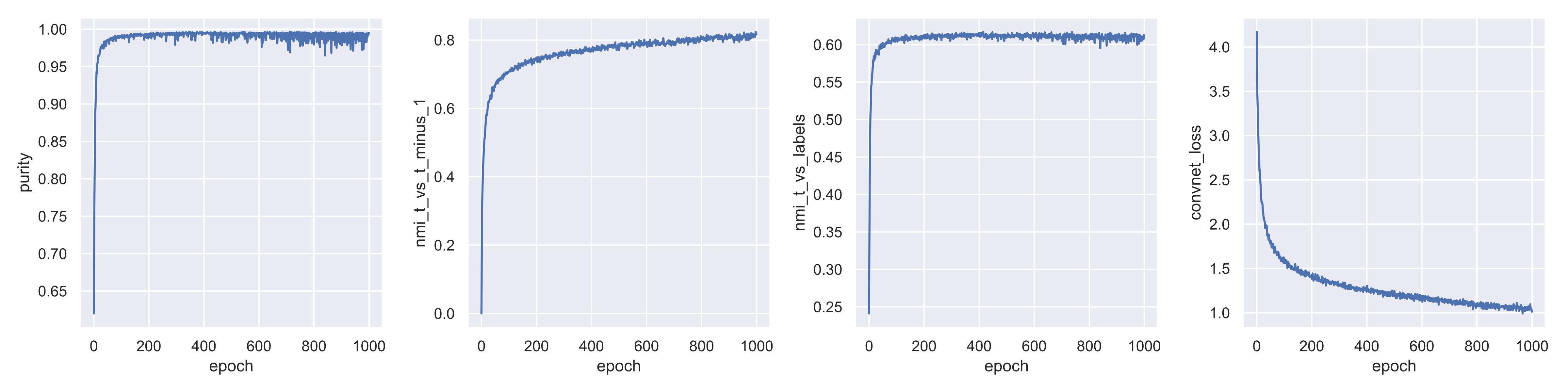}
\caption{Training stability and metrics for 1000 epochs} \label{metrics_stability_2}
\end{figure}

Additionally, we want to extend the discussion about deep clustering at \cite{caron2018deepcluster} to medical imaging in a realistic scenario. One major challenge in deep clustering is the lack of a stopping criterion. Supervised training with labelled data as a stopping criterion could be utilized but this usually requires the prior knowledge of classes, which may not be possible to have beforehand at an unstructured hospital database.
Pre-defined threshold-based methods on evaluation metrics, e.g. NMI and purity from adjacent epochs \cite{wang2017unsupervised}, could be another option but their robustness has yet to be proven. This is why it is important to investigate whether the training diverges. For this aim, we trained the dataset (iii) with 1000 epochs to observe the training stability. As we can see from Fig~\ref{metrics_stability_2}, although we observed some fluctuations in metrics from time to time, they were stable throughout the training, which is similar to the observation at \cite{caron2018deepcluster}.

Lastly, we observed that changes in NMI and CNN loss could indicate changes in clustering quality. Normally, we expect to see a steady increase in NMI and a steady decrease in CNN loss during the training. A sudden decrease in NMI and/or a sudden increase in CNN loss may be a sign of worse clusters generated. However, steady decrease in CNN loss does not necessarily mean better cluster purity. Therefore, we think that it is beneficial to closely monitor not one but all metrics for unusual changes as well as to consider other metrics of clustering.

\section{Conclusion}
In this work, we propose an unsupervised deep clustering approach with end-to-end training to automatically categorize large-scale medical images without using any labels. We have demonstrated that our method is able to generate highly pure clusters (above 0.99) under both balanced and imbalanced class distributions. In future work, expanding the evaluation, adapting deep clustering approaches to other clinical tasks and improving their robustness and generalizability are some of interesting avenues that could be explored. 

\section{Acknowledgement}
This work is supported by the UK Research and Innovation London Medical Imaging and Artificial Intelligence Centre for Value Based Healthcare. This research has been conducted using the UK Biobank Resource under Application Number 12579.

\bibliographystyle{splncs04}
\bibliography{references}


\newpage
\section{Supplementary Material}

\setlength{\tabcolsep}{0.5em} 
\begin{table}[ht]
\centering
\caption{Class distributions for all datasets}
\begin{tabular}{lccc}
  &  \textbf{dataset (i)}  &  \textbf{dataset (ii)}  &  \textbf{dataset (iii)}   \\ 
\hline
AO &  0  &  7859  &   982  \\
FLOW &  0  &  7782  &  971   \\
FLOW MAG &  0  &  7782  &  971   \\
FLOW PHA &  0  &  7782  &  971   \\
LA (2 ch) &  15868  &  7931  &   990  \\
LA (3 ch) &  15889  & 7943   &   992  \\
LA (4 ch) &  15880  &  7937  &  990   \\
LVOT &  0  &  7831  &   979  \\
SA &  0  &  83372  &   10339   \\
SHMOLLI &  0  &  7565  &   944  \\
SHMOLLI FITPAR &  0  &  7561  &  944   \\
SHMOLLI T1MAP &  0  &  7560  &  944   \\
CINE TAG &  0  &  23367  &  2926  
\end{tabular}
\end{table}

\begin{figure}
\centering
\includegraphics[width=\textwidth]{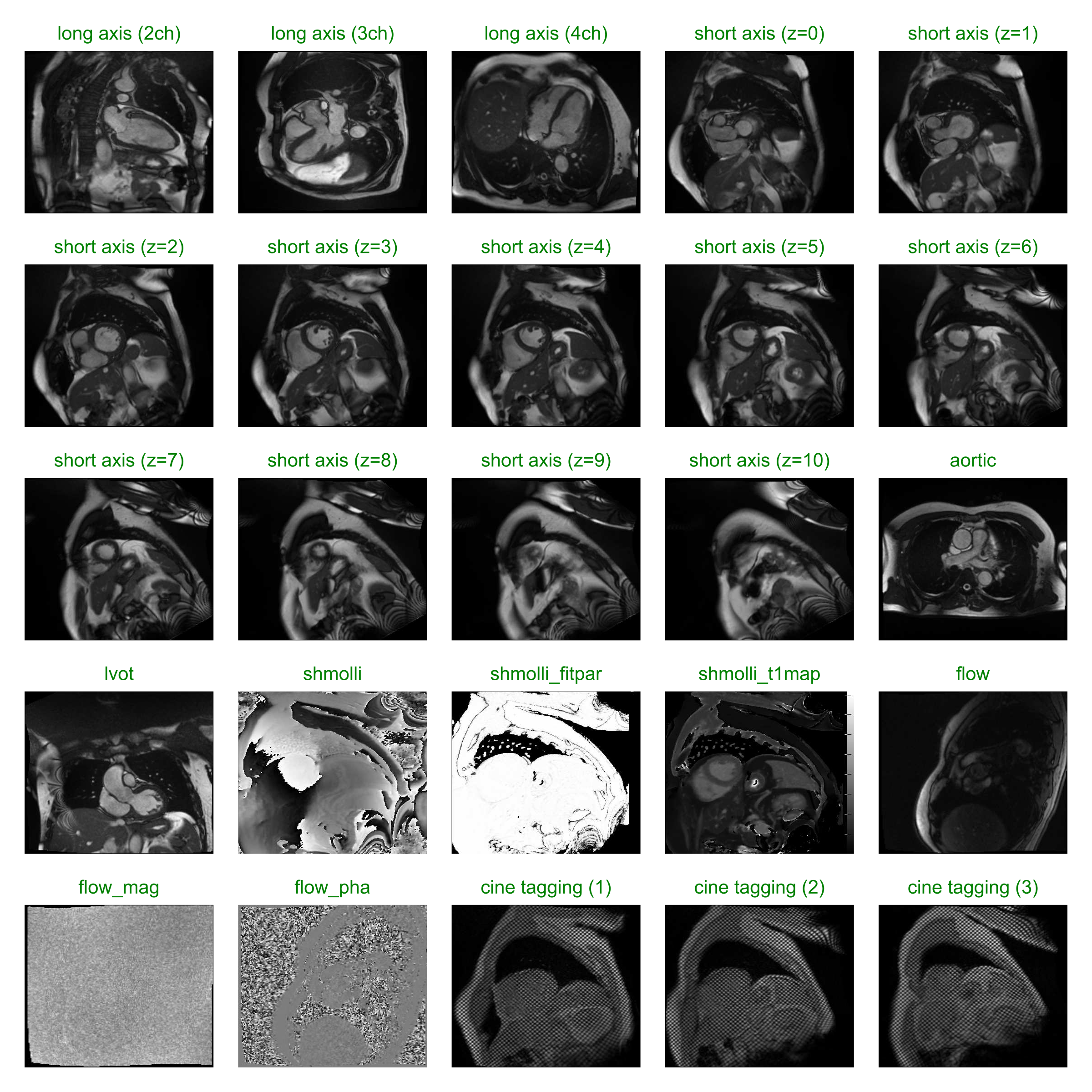}
\caption{Example cardiac images from the datasets} \label{examples}
\end{figure}

\end{document}